\documentstyle[preprint,aps]{revtex}
\begin{document}
\draft
\title{Shell structure and electron-electron interaction in self-assembled InAs quantum dots}
\author{M. Fricke, A. Lorke, J. P. Kotthaus}
\address{Sektion Physik, LMU M\"unchen, 
Geschwister-Scholl-Platz 1, 80539 M\"unchen, Germany}
\author{G. Medeiros-Ribeiro and P. M. Petroff}
\address{Materials Department and QUEST, University of California, Santa 
Barbara, CA 93117}
\date{\today}
\maketitle

\begin{abstract}
Using far-infrared spectroscopy, we investigate the excitations of self-organized InAs quantum dots as a function of the electron number per dot, $1\leq n_e\leq 6$, which is monitored {\it in situ} by capacitance spectroscopy. Whereas the well-known two-mode spectrum is observed when the lowest  ($s$-) states are filled, we find a rich excitation spectrum for $n_e\geq 3$, which reflects the importance of electron-electron interaction in the present, strongly non-parabolic confining potential. 
From capacitance spectroscopy we find that the electronic shell structure in our dots gives rise to a distinct pattern in the charging energies which strongly deviates from the monotonic behavior of the Coulomb blockade found in mesoscopic or metallic structures.
\end{abstract}

The influence of the electron-electron interaction on the static and dynamic transport properties of quantum dots has been a topic of continuing interest ever since these man-made, few-electron systems became available. On the one hand, the Coulomb repulsion between the electrons has dramatic effects on the conductance through a quantum dot, as evidenced, e.g. in the so-called Coulomb blockade.\cite{CoulombB} On the other hand, the generalized Kohn's theorem\cite{Maksym,Yip} states that in dots with a sufficiently smooth confining potential the high-frequency dipole excitations are unaffected by electron-electron interactions. 

Recently, heterostructures have been made possible \cite{Drexler} that combine the advantages of single-electron capacitance spectroscopy \cite{Hansen,Ashoori} with those of self-assembled quantum dots.\cite{Leonard} These samples offer the unique opportunity to observe single-electron charging in large-scale quantum dot arrays which make it possible to study simultaneously the dots' dynamic response in the far-infrared.

In the present Letter we study the far-infrared (FIR) absorption of self-assembled, nm-size InAs quantum dots, where the number of occupied electron states can be finely tuned and monitored by single-electron capacitance spectroscopy. This way, the absorption is investigated for $n_e=1$ to 6 electrons per dot. The combination of capacitance and FIR spectroscopy enables us to extract detailed information about the quantum and charging energies of the $s$- and $p$-states in InAs/GaAs quantum dots. A comparison is made with recent calculations\cite{Hawrylak,Peeters} of the many-particle states and transition energies of self-assembled quantum dots.

The samples are grown by molecular-beam epitaxy using the now well-established Stranski-Krastanow growth procedure.\cite{Drexler,Leonard,Goldstein,Medeiros,Madhukar,Bimberg,Marzin} The InAs dots are embedded into a MISFET (metal-insulator-semiconductor-field-effect-transistor)-type GaAs/AlGaAs heterostructure, as described in Refs.\ \onlinecite{Drexler,Medeiros}. In the present sample the dots are located 150 nm below the surface and 25 nm above the highly doped back contact layer. From atomic force micrographs of similarly grown samples\cite{Leonard} we estimate the InAs dots to be approximately 20 nm in diameter and 7 nm in height.

The sample is provided with Ohmic contacts and a semi-transparent gate. It is  then mounted in a liquid He cryostat in the center of a superconducting solenoid which allows magnetic fields of up to $B=15$ T to be applied perpendicular to the sample surface. The capacitance-voltage (CV) measurements are carried out using a standard lock-in technique with a modulation frequency of 278 Hz and an amplitude of $\Delta U=2$ mV. The FIR transmission is recorded by rapid-scan Fourier-transform spectroscopy in the energy range between 5 and 100 meV. In order to eliminate unwanted spectral features from the optical components in the set-up, the sample spectrum is normalized by a reference spectrum taken at a gate voltage were the dots are void of electrons.

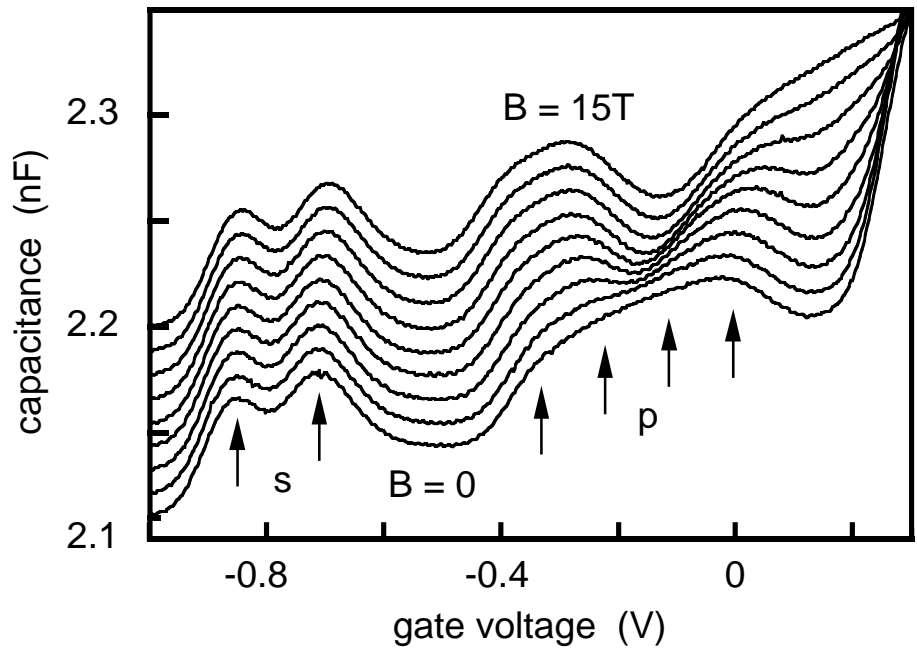
\begin{figure}
\caption{Capacitance-voltage traces of self-assembled InAs quantum dots for magnetic fields $B=$ 0, 2, 4, 6, 8, 10, 12, 13.5, and 15 T. Single-electron charging of the $s$-state can be observed around $V_g=-0.8$ V. The charging peaks of the $p$-state ($V_g\approx-0.2$ V) have merged into one broad structure which shows the characteristic Zeeman-splitting at high magnetic fields. The curves have been offset for clarity.}
\label{FigCV}
\end{figure}

Figure \ref{FigCV} shows typical CV traces of our samples in magnetic fields between $B=0$ and 15 T. At low gate voltages, $V_g \leq -1$ V, the signal is given by the geometric capacitance between the gate and the back contact. As the bias is increased, the states in the layers between the back contact and the gate become occupied and their density of states gives rise to distinct features in the CV trace.\cite{Drexler,Medeiros} The double peak at $V_g \approx -800$ mV corresponds to the filling of the lowest ($s$-) state of the quantum dots. This assignment can be verified by the magnetic field dependence and by the fact that no dot-induced FIR absorption can be observed for $V_g \leq -900$ mV. The $s$-state is doubly spin-degenerate, and the difference in gate voltage $\Delta V_g^{12}=147$ mV between loading of the first and the second electron in Fig.\ \ref{FigCV} is a direct consequence of the Coulomb blockade caused by the electron-electron interaction in the dots. From $\Delta V_g^{12}$ we can determine the charging energy $E_{e-e}^{12}$ between the singly and doubly occupied dot states. Taking into account the image charge of the metallic back contact we calculate the relation between $\Delta V_g$ and $E_{e-e}$ as

\begin{equation}
E_{e-e}=e\frac{t_b}{t_{tot}}\Delta V_g + \frac{e^2}{8\pi \varepsilon \varepsilon_0 t_b}
\label{eqEee}\end{equation}

Here, $t_b$ is the distance between the back contact and the dots, $t_{tot}$ is the distance between the back contact and the gate. Because of the low dot density in our samples, we have neglected the screening of the electric field between the back contact and the gate by the dots. For $t_b,\ t_{tot}\rightarrow\infty$, Eq. \ref{eqEee} reduces to the lever-arm argument given, e.g. in Refs.\ \onlinecite{Drexler,Medeiros}. With $\Delta V_g^{12}=147$ mV and $t_b/t_{tot}=1/7$, Eq. \ref{eqEee} gives an $s$-state charging energy of 23.3 meV. 

In our oblate, nearly circular dots\cite{Leonard} the second ($p$-) shell is fourfold degenerate and, because of Coulomb repulsion, we expect four peaks at higher gate voltages, when the $p$-state is gradually filling. In Fig.\ \ref{FigCV}, however, these peaks are not resolved but have merged to form one broad maximum around $V_g= -200$ mV. At higher gate biases, $V_g>+200$ mV, the wetting-layer becomes occupied, as reflected by the strong increase of the capacitance, so that charging of higher dot states can no longer be observed.

Figure \ref{FigCV} shows that, as expected, the $s$-state is only little affected by magnetic fields up to $B=15$ T, whereas the $p$-state shows the characteristic Zeeman splitting,\cite{Drexler,Hansen} with two states increasing with increasing field and two decreasing. The spin splitting at these magnetic fields is too small to give a significant contribution to the charging characteristics of the dot states.

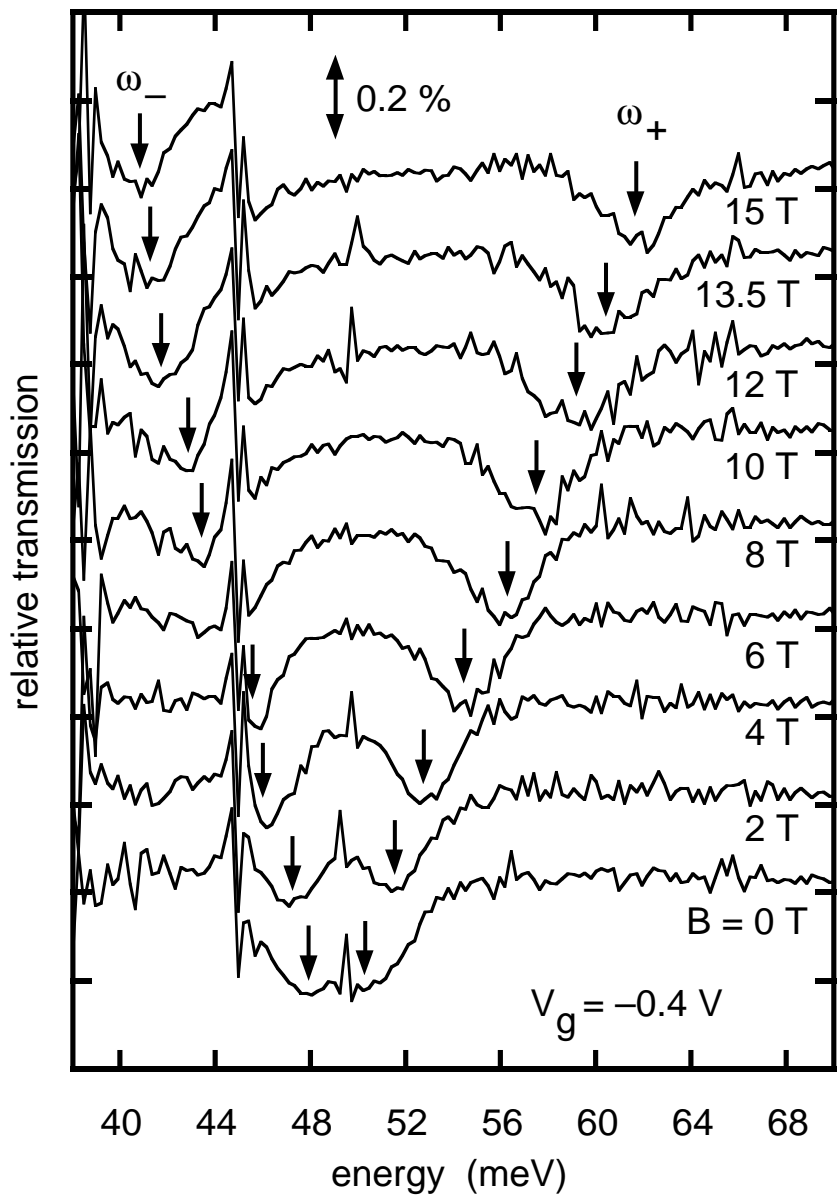
\begin{figure}
\caption{Normalized transmission of doubly occupied dots in magnetic fields between 0 and 15 T. The typical two-mode spectrum of quantum dots in a magnetic field can be observed. The sharp feature at $\approx$ 45 meV originates from an electronic interaction with an interface phonon.\protect\cite{Batke} Traces have been offset for clarity.}
\label{FigTransm-0.4V}
\end{figure}

Figure \ref{FigTransm-0.4V} displays the FIR transmission of the sample at magnetic fields between 0 and 15 T and a gate voltage of $-400$ mV. At this bias the $s$-state is completely filled whereas the $p$-state is still empty (see Fig.\ref{FigCV}). The well-known two-mode spectrum of quantum dots\cite{HansenWB} can be observed with one resonance, $\omega_+$, increasing in magnetic field and the other, $\omega_-$, decreasing. Because of a signal-to-noise ratio $>10^3$, we are able to detect the $\omega_-$-mode, which, to our knowledge, has not been observed before in self-assembled InAs dots. Furthermore, we can determine a minute splitting of the resonance at $B=0$ of approximately 2 meV. Polarization-dependent measurements show that at $B=0$ the upper (lower) mode is excited by FIR radiation polarized along the $[110]$ ([1\={1}0]) direction. This can be explained by a slight elongation of the dots of only $\approx 5$ {\AA} along [1\={1}0], which might be a remainder of the extreme anisotropy of InAs islands on GaAs for submonolayer coverage.\cite{BresslerHill} A slight elongation of InAs dots has been suggested, e.g., by Nabetani {\it et al.} \cite{Nabetani}, however, to our knowledge, the data in Fig.\ \ref{FigTransm-0.4V} represents the first experimental evidence for the associated breaking of the symmetry in the electron states. Alternatively, a lifting of the symmetry by  piezoelectric effects, as proposed by Grundmann {\it et al.} \cite{Grundmann} could account for the observed splitting. 

From the magnetic field dependence of the $\omega_+$- and $\omega_-$-modes we can deduce an effective cyclotron mass of $m^*=0.080\pm0.003 m_e$. This value is surprisingly high, considering that the band edge masses of InAs and GaAs are $0.023 m_e$ and $0.067 m_e$, respectively. Possible explanations for such a high effective mass could be strain effects, non-parabolicity\cite{nonparab}, or the fact that a large fraction of the dot wave functions spread into the GaAs\cite{Peeters} {\em below} the conduction band edge.\cite{Roessler} An evaluation of the $p$-state splitting observed by capacitance spectroscopy (cf. Fig.\ \ref{FigCV}) leads to a somewhat lower, however less accurate, mass of $m^*=0.066\pm0.015 m_e$. 

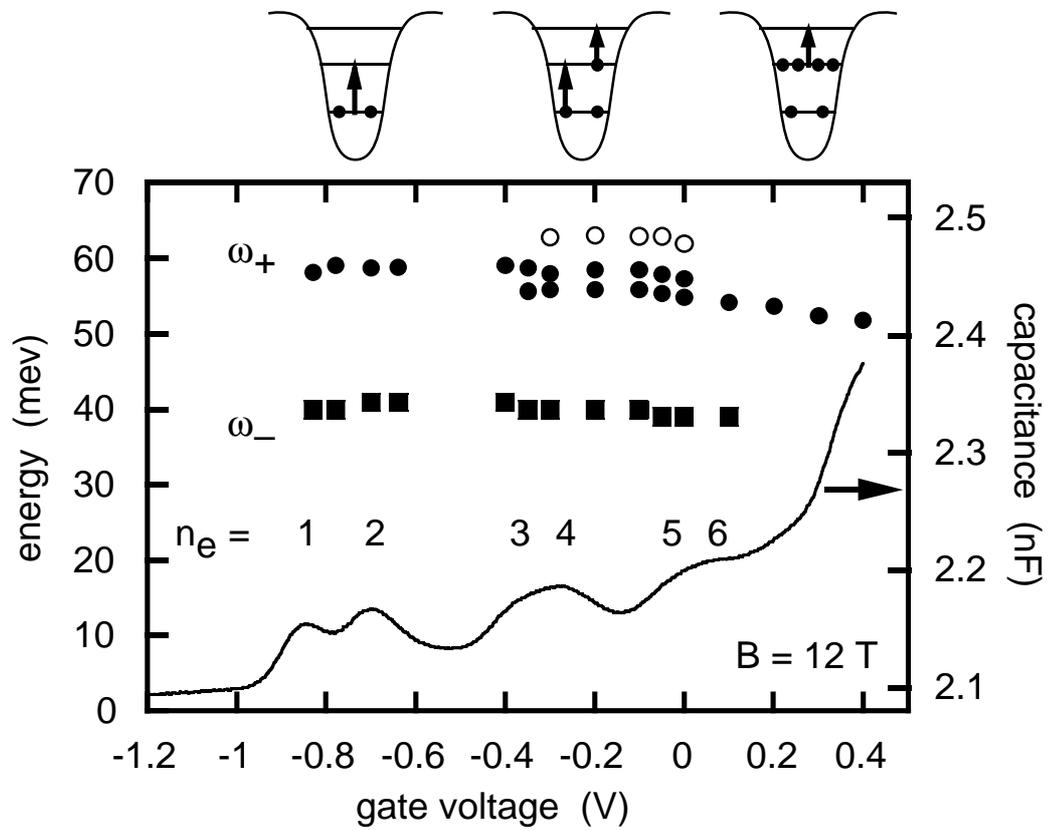
\begin{figure}
\caption{Far-infrared transition energies $\omega_+$ and $\omega_-$ of self-assembled quantum dots at $B=12$ T as a function of gate voltage (left scale). The number of occupied electron states, $n_e$, as deduced from the CV trace (solid line, right scale) is indicated.}
\label{figFIR12T}
\end{figure}

As mentioned above, the simultaneously taken capacitance spectra enable us to examine the FIR-response of the dots as a function the electron occupation number. Figure \ref{figFIR12T} displays the position of the FIR-resonances as a function of the gate voltage at $B=12$ T. Also shown are the CV trace at this magnetic field and the electron number per dot $n_e$ deduced from it. For $n_e\leq2$, i.e. when only the $s$-state is occupied, we observe only two resonances. The energetic positions of these resonances are independent of $n_e$ over a large range of gate biases. This situation changes abruptly, as soon as the third electron is loaded into the dots and the $p$-state becomes occupied. Then the $\omega_+$-mode splits up into up to three resonances. These resonances can be observed in the bias regime $-500$ mV $\leq V_g\leq +200$ mV where the $p$-state is (partly) occupied. 

All these observations are in good qualitative agreement with recent calculations by Wojs and Hawrylak,\cite{Hawrylak} who have calculated the electronic states and FIR excitation energies of self-assembled InGaAs dots. From our experimental observations and a comparison with theoretical models\cite{Hawrylak,Peeters,Pfannkuche} we draw the following conclusions. For $n_e\leq2$ the dynamic behavior of the dots is very similar to that of a parabolically confined system. Only two modes are observed which follow the expected dispersion of a harmonic oscillator in a magnetic field.\cite{Maksym,Yip} Furthermore, the energetic position of these modes are independent of $n_e$, in agreement with the generalized Kohn theorem. For $n_e\geq 3$, new transitions become possible which demonstrate the deviation from a purely parabolic potential present in our dots. The inset in Fig.\ \ref{figFIR12T} illustrates this from a simple, single-particle point of view. As long as the $p$-state is empty, only $s\rightarrow p$ transitions are possible which resemble those of a purely parabolic dot. As soon as the $p$-state becomes occupied, $p\rightarrow d$ transitions can be excited, which, due to a flattening out of the confining potential at higher energies, are expected to have a lower energy than the $s\rightarrow p$ transitions. We therefore identify the lowest $\omega_+$-mode as originating from $p\rightarrow d$ transitions. The next higher $\omega_+$-mode is then attributed to an $s\rightarrow p$ transition. This agrees with the fact that its energy coincides with that of the $\omega_+$-mode at lower gate voltages and that this mode vanishes when the $p$-state is completely filled and $s\rightarrow p$ transitions are no longer possible. However, a single-particle picture can only be a rough approximation to the behavior of our dots for $n_e\geq3$ where electron-electron interactions already play an important role. This is clearly demonstrated by the appearance of a third $\omega_+$-mode (open symbols) which is only present when the $p$-state is partially filled. The single particle picture, above, cannot account for this resonance. A calculation of the dot excitations which takes into account electron-electron interactions, however, indeed predicts the occurrence of up to three $\omega_+$-modes for a partially filled $p$-state. \cite{Hawrylak} 

Further information on the electron-electron interaction in InAs quantum dots can be obtained by a careful analysis of the capacitance data. To obtain an approximate position of the $p$-state charging energies, we deconvolve the broad $p$-state capacitance signal (cf. Fig.\ \ref{FigCV}) into four equidistant bell curves of equal width.\cite{Barbara} A charging energy between the $p$-states of approximately 18 meV is obtained, which is somewhat smaller than the Coulomb blockade in the $s$-state of 23 meV mentioned above. This is in qualitative agreement with the charging energy of a Coulomb island $E_C=e^2/C$ where the island's capacitance $C$ is proportional to its diameter, which, in turn, is expected to increase with increasing electron number in semiconductor quantum dots. However, a distinct deviation from this behavior can be identified, when the Coulomb blockade between the second $s$-state and the first $p$-state is considered. Here, we derive a charging energy of only $\approx$ 7 meV when we subtract the "bare" quantization energy of $\hbar \omega_0\approx$ 50 meV from the energy difference determined by capacitance spectroscopy. In a simple, perturbative approach, this low charging energy directly reflects the small overlap between the $s$- and the $p$-state as compared to two $s$- or two $p$-states and shows the influence of the electronic shell structure on the Coulomb repulsion. Again, a more complete, many-electron treatment leads to the same qualitative behavior:\cite{Hawrylak} For InGaAs dots with a quantization energy of $\hbar\omega_0\approx$ 50 meV, Wojs and Hawrylak obtain charging energies of 30 meV, 11 meV, and an average 20 meV for the $s$-$s$, $s$-$p$, and $p$-$p$ charging energies, respectively.

As pointed out by Merkt {\it et al.} \cite{MerktundCo}, the Coulomb energy of two electrons in a parabolic dot $E_{e-e}^{12}$ and the characteristic length of the ground state $\ell$ are related by $E_{e-e}^{12}=e^2/(4\pi \varepsilon\varepsilon_0 \ell)$. For the present dots this leads to $\ell=4.9$ nm. On the other hand, this characteristic length can be determined using the  effective mass and the excitation frequency at $B=0$, $\omega_0$, obtained by FIR spectroscopy (neglecting the 5\% splitting). We find $\ell=\sqrt{\hbar/(m^* \omega_0)}=4.4$ nm in agreement with the above value, which nicely demonstrates the applicability of these models for the present few-electron quantum systems.

In summary, we have probed by far-infrared and capacitance spectroscopy the $n$-electron ground state and excitations of self-assembled InAs quantum dots.  As predicted by recent theoretical models, we find a rich excitation spectrum when both $s$- and $p$-states are (partly) occupied. This enables us to distinguish, e.g. quantum dot "Helium" from quantum dot "Lithium" by their characteristic excitation frequencies. This has been a long-standing goal in the field of semiconductor nanostructures, which had so far been obstructed by the generalized Kohn theorem. \cite{Heitmann}

We would like to thank V. Dolgopolov and W. Hansen for stimulating discussions and P. Hawrylak and F. Peeters for making their results known to us prior to publication. Financial support through the Deutsche Forschungsgemeinschaft and through QUEST, a NSF Science and Technology Center, under Grant No. DMR 20007, is gratefully acknowledged.


\end{document}